\documentclass[
reprint,
showpacs,
amsmath,
amssymb,
aps,
]{revtex4-1}

\usepackage{tikz}
\usepackage{pgffor}
\usepackage{verbatim}
\usepackage{pstricks}
\usepackage{pst-3dplot}
\usepackage[colorlinks, breaklinks=true, linkcolor=purple, citecolor=purple, linktocpage=true]{hyperref}
\usepackage{color}
\usepackage{graphicx}
\usepackage{dcolumn}
\usepackage{bm}

\usepackage{txfonts}

\begin{document}

\title{Electric-field-induced spin resonance in antiferromagnetic insulators:\\ Inverse process of the dynamical chiral magnetic effect}

\author{Akihiko Sekine}
\author{Takahiro Chiba}
\affiliation{Institute for Materials Research, Tohoku University, Sendai 980-8577, Japan}

\date{\today}

\begin{abstract}
We propose a realization of the electric-field-induced antiferromagnetic resonance.
We consider three-dimensional antiferromagnetic insulators with spin-orbit coupling characterized by the existence of a topological term called the $\theta$ term.
By solving the Landau-Lifshitz-Gilbert equation in the presence of the $\theta$ term, we show that, in contrast to conventional methods using ac magnetic fields, the antiferromagnetic resonance state is realized by ac electric fields along with static magnetic fields.
This mechanism can be understood as the inverse process of the dynamical chiral magnetic effect, an alternating current generation by magnetic fields.
In other words, we propose a way to electrically induce the dynamical axion field in condensed matter.
We discuss a possible experiment to observe our proposal, which utilizes the spin pumping from the antiferromagnetic insulator into a heavy metal contact.
\end{abstract}


\maketitle

{\it Introduction.---}
Possible applications of materials in which the low-energy effective models are described by relativistic Dirac fermions have been studied intensively and extensively.
The representative examples of Dirac fermion systems are graphene \cite{Castro-Neto2009} and topological insulators \cite{Hasan2010,Qi2011,Ando2013}.
Studies on topological phases of matter are now extended to various gapless topological phases such as Weyl semimetals \cite{Murakami2007,Wan2011,Lv2015,Yang2015,Xu2015} and Dirac semimetals \cite{Wang2012,Wang2013,Liu2014,Liu2014a}.
Unlike graphene, topological phases are realized in spin-orbit coupled systems.
It has been revealed that strong spin-orbit coupling (SOC) is a key element in unconventional phenomena.
For example, the spin-momentum locking, which is realized in the metallic surface states of three-dimensional (3D) topological insulators, arises as a consequence of SOC and results in a persistent pure spin current on the surface that is robust against disorder \cite{Hasan2010,Qi2011,Ando2013}.
Recently, possible ways to manipulate and utilize such topological surface states have been investigated experimentally in spintronics \cite{Mellnik2014,Li2014,Fan2014,Shiomi2014}.
Another interesting example is the chiral magnetic effect, which was originally proposed in high-energy theory as a direct current generation by static magnetic fields \cite{Fukushima2008}.
Its possibility has been investigated theoretically in Weyl semimetals \cite{Zyuzin2012,Vazifeh2013,Goswami2013,Burkov2015,Chang2015,Buividovich2015}, and the dynamical realization of the chiral magnetic effect, i.e., an alternating current generation by magnetic fields, has also been proposed in spin-orbit coupled insulators \cite{Sekine2016}.

One of the most important themes in spintronics is the generation and control of spin currents.
Spin pumping is a powerful technique to generate a pure spin current \cite{Tserkovnyak2005,Saitoh2006,Ando2011}.
While a magnetization is precessing, the spin angular momentum in a magnet is injected across the interface into a neighboring material through the exchange interaction.
The spin current injected into a heavy metal (HM) such as Pt can be detected electrically via the inverse spin Hall effect \cite{Saitoh2006,Hoffmann2013,Sinova2015}.
So far, ferromagnets have been used as a source material in which magnetization precession is caused for spin pumping.
On the other hand, unlike ferromagnets, antiferromagnets had not been considered to be of practical use due to zero net magnetization.
However, recent studies in spintronics is now extended to active use of antiferromagnets \cite{MacDonald2011,Cheng2014,Wang2014,Zhang2014,Jungwirth2015}.
It has been suggested that antiferromagnets can complement or replace ferromagnets as active elements of a memory \cite{Marti2014} or logic device \cite{Wadley2016}, e.g., because antiferromagnets do not generate unwanted stray fields.

In this paper, we propose a realization of the electrically driven antiferromagnetic (AF) resonance in 3D AF insulators with SOC.
This is in sharp contrast to conventional methods using magnetic fields.
In spintronics, electrical manipulation of magnetism is one of the most important subjects in the pursuit of energy-saving and higher-density information storage.
However, preceding studies have been based on electric-current-induced methods that require such high-density currents as $\sim 10^{10}\ \mathrm{A/m^2}$ \cite{Brataas2012}.
Namely, there is a large energy dissipation due to Joule heat.
In contrast, since the system we consider is insulating, there is no energy dissipation due to Joule heat in the presence of electric fields.
Moreover, whereas the electric-field-induced ferromagnetic resonance has been realized experimentally in a ferromagnetic metal \cite{Nozaki2012}, the electric-field-induced AF resonance has not yet been realized.
The key ingredient for its realization in our study is a topological term called the $\theta$ term which arises due to strong SOC.
The existence of the $\theta$ term leads to a coupling of electric fields and the N\'{e}el field.
By solving the Landau-Lifshitz-Gilbert equation in the presence of the $\theta$ term, we show that the AF resonance state is realized by ac electric fields.
We also show that the resonance state can be detected as an usual spin-pumping-induced voltage signal.
We argue that the mechanism of the electric-field-induced AF resonance in this study can be understood as the inverse process of the dynamical chiral magnetic effect \cite{Sekine2016}.

%%%%%%%%%%%
{\it Low-energy effective model.---}
We study a class of 3D AF insulators that can be realized in systems with electron correlations and SOC, such as $5d$ transition metal oxides \cite{Witczak-Krempa2014,Rau2015}.
As a theoretical model, we adopt a tight-binding model called the Fu-Kane-Mele-Hubbard model on a diamond lattice \cite{Sekine2014,Sekine2016}, in which the nearest-neighbor electron hopping, spin-dependent next-nearest-neighbor electron hopping (i.e., SOC), and on-site electron-electron repulsive interaction are taken into account.
In this model, an AF insulator phase develops when on-site interactions are strong.
Here, the mean-field AF order parameter is parameterized between the two sublattices as $(n_0\sin\theta\cos\varphi,n_0\sin\theta\sin\varphi,n_0\cos\theta)$, with $n_0$ the magnitude of the order parameter, and angles $\theta$ and $\varphi$ obtained from the coordinate of a diamond lattice \cite{Sekine2014}.

The mean-field low-energy effective action of the system in the presence of an external electromagnetic field is given by a massive Dirac fermion model of the form \cite{Sekine2014,Sekine2016}
\begin{align}
\begin{split}
S_{\rm AF}=\int dtd^3 r\sum_{f=1,2,3}\bar{\psi}_f\left[i\gamma^\mu D_\mu-M_0+i\gamma^5M_{5f}\right]\psi_f,\label{Eff-Action}
\end{split}
\end{align}
where $t$ is real time, $\psi_f(\bm{r},t)$ is a four-component spinor in the basis of the sublattice degrees of freedom of the diamond lattice and electrons' spin degrees of freedom, $\bar{\psi}_f=\psi_f^\dagger\gamma^0$, $D_\mu=\partial_\mu+ieA_\mu$ with $A_\mu$ an electromagnetic potential.
The $4\times 4$ matrices $\gamma^\mu$ with $\gamma^5=i\gamma^0\gamma^1\gamma^2\gamma^3$ are the Dirac gamma matrices.
The subscript $f$ denotes the valley degrees of freedom.
$M_0$ is a mass term (band gap) induced by strong SOC, and preserves time-reversal and inversion symmetries.
$i\gamma^5M_{5f}$ is a mean-field mass term induced by the AF ordering of itinerant electrons, and breaks both time-reversal and inversion symmetries.
$M_{5f}$ are given explicitly by $M_{5,1}=Un_0n_1$, $M_{5,2}=Un_0n_2$, and $M_{5,3}=Un_0n_3$, where $U$ is the strength of on-site electron-electron repulsive interactions, and $(n_1,n_2,n_3)=(\sin\theta\cos\varphi,\sin\theta\sin\varphi,\cos\theta)$ is the N\'{e}el field.
Effective actions similar to Eq. (\ref{Eff-Action}) have been obtained in the AF insulator phases of the magnetically doped Bi$_2$Se$_3$ family \cite{Li2010} and transition metal oxides with the corundum structure \cite{Wang2011}.

Integrating out the fermionic field $\psi_f$, we obtain the effective action in terms of the N\'{e}el field and an electromagnetic field up to the relevant lowest order in $n_f$ as $S_{\rm AF}=S_0+S_\theta$ \cite{Sekine2016}.
Here, $S_0$ is the action of the N\'{e}el field (i.e., the nonlinear sigma model \cite{Haldane1983})
\begin{align}
\begin{split}
S_0=\frac{1}{g}\int dtd^3r \left[(\partial_\mu \bm{n})\cdot(\partial^\mu \bm{n})+\Delta_0^2\bm{n}^2\right],\label{Action-NLSigma}
\end{split}
\end{align}
and $S_\theta$ is a topological term called the $\theta$ term \cite{Qi2008}
\begin{align}
\begin{split}
S_\theta=\frac{e^2}{2\pi h}\int dtd^3 r \theta(\bm{r},t) \bm{E}\cdot\bm{B},\label{theta-term}
\end{split}
\end{align}
where $g$ is a constant, $\Delta_0$ is the spin-wave gap, $\bm{E}$ ($\bm{B}$) is an external electric (magnetic) field, and $\theta(\bm{r},t)=\frac{\pi}{2}[1+\mathrm{sgn}(M_0)]-(Un_0/M_0)\sum_{f=1,2,3}n_f(\bm{r},t)$.
Here, $\theta(\bm{r},t)$ is known as the {\it dynamical axion field} \cite{Li2010}.
The $\theta$ term results in the topological magnetoelectric effect in the bulk such that $\bm{P}=(e^2/2\pi h)\theta\bm{B}$ and $\bm{M}=(e^2/2\pi h)\theta\bm{E}$ with $\bm{P}$ the electric polarization and $\bm{B}$ the magnetization \cite{Hasan2010,Qi2011,Ando2013,Qi2008}.
In the presence of time-reversal symmetry, $\theta=\pi$ (mod $2\pi$) in 3D topological insulators and $\theta=0$ in normal insulators.
However, the value of $\theta$ can be arbitrary in such systems with broken time-reversal and inversion symmetries as AF insulators described by Eq. (\ref{Eff-Action}) \cite{comment-A}.

Let us implement a little more realistic condition in the above model.
We take into account a small net magnetization $\bm{m}$ satisfying the constraint $\bm{n}\cdot\bm{m}=0$ with $|\bm{n}|=1$ and $|\bm{m}|\ll 1$.
Furthermore, we assume the case of AF insulators with easy-axis anisotropy.
Note that the magnetic anisotropy direction in Eq. (\ref{Action-NLSigma}) cannot be determined from Eq. (\ref{Eff-Action}), since the N\'{e}el field is isotropic in Eq. (\ref{Eff-Action}).
Then a modification of Eq. (\ref{Action-NLSigma}) gives the free energy of the trivial part as \cite{LL-book,Hals2011}
\begin{align}
\begin{split}
F_0=\int d^3r \left[\frac{a}{2}\bm{m}^2+\frac{A}{2}\sum_{i=x,y,z}(\partial_i\bm{n})^2-\frac{K}{2}n_z^2-\bm{H}\cdot\bm{m}\right],
\end{split}
\end{align}
where $a$ and $A$ are the homogeneous and inhomogeneous exchange constants, respectively, and $K$ is the easy-axis anisotropy along the $z$ direction.
The fourth term is the Zeeman coupling with $\bm{H}=g\mu_B\bm{B}$ being an external magnetic field.
In the following, we define a laboratory frame in which the $z$ direction is set to be the easy-axis direction.
On the other hand, the free energy of the topological part is given by
\begin{align}
\begin{split}
F_\theta=-\frac{e^2}{2\pi h}\frac{\sqrt{3}Un_0}{M_0}\int d^3r (\bm{n}\cdot\bm{e}_{[111]})\bm{E}\cdot\bm{B},\label{F-theta}
\end{split}
\end{align}
where we have used the fact that $\sum_{f=1,2,3}n_f=(n_1\bm{e}_1+n_2\bm{e}_2+n_3\bm{e}_3)\cdot(\bm{e}_1+\bm{e}_2+\bm{e}_3)=\sqrt{3}\bm{n}\cdot\bm{e}_{[111]}$ with $\bm{e}_{[111]}$ being the unit vector along the [111] direction of the original diamond lattice in the Fu-Kane-Mele-Hubbard model.

\begin{figure}[!t]
\centering
\includegraphics[width=\columnwidth]{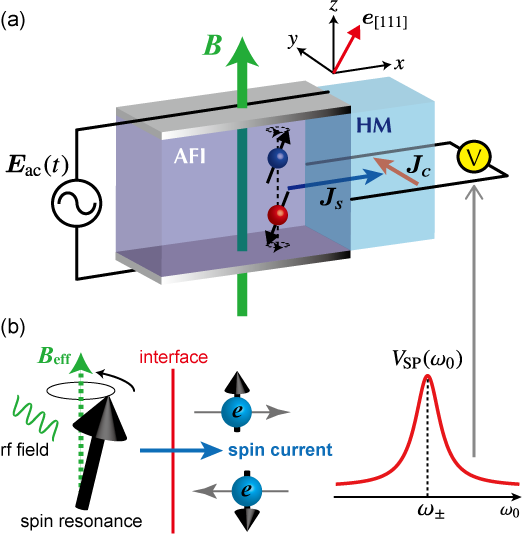}
\caption{(a) Schematic figure of a possible experimental setup to observe the electric-field-induced AF resonance in our system.
The direction of the AF order (i.e., the easy axis) is defined as the $z$ direction.
$\bm{e}_{[111]}$ is the unit vector along the [111] direction of the diamond lattice. 
An ac electric field $\bm{E}_{\mathrm{ac}}(t)=E_{\mathrm{ac}}e^{i\omega_0 t}\bm{e}_z$ is applied to the parallel-plate capacitor, and a static magnetic field $\bm{B}=B\bm{e}_z$ is also applied to the AF insulator.
The pumped spin current $\bm{J}_{s}$ into the attached HM such as Pt can be detected through the inverse spin Hall effect as a direct current $\bm{J}_{c}$ (i.e., the voltage $V_{\rm SP}$) across the transverse direction.
(b) Schematic illustration of the spin pumping mechanism.
Spin resonance injects a pure spin current into the attached layer via the exchange interaction at the interface.
}\label{Fig1}
\end{figure}

%%%%%%%%%%%
{\it Electric-field-induced antiferromagnetic resonance.---}
In order to realize and detect the electric-field-induced AF resonance, we consider the AF insulator/HM bilayer system in the presence of an ac electric field $\bm{E}_{\mathrm{ac}}(t)=E_{\mathrm{ac}}e^{i\omega_{0}t}\bm{e}_z$ and a static magnetic field $\bm{H}=g\mu_B B\bm{e}_z$ ($\bm{B}=B\bm{e}_z$) with $B$ being much weaker than both the AF exchange coupling and easy-axial anisotropy.
Here $\bm{e}_z$ is the unit vector parallel to the easy axis of the AF order.
A schematic figure of our setup is shown in Fig.~\ref{Fig1}(a).
The essential point is the coupling of the N\'{e}el field and an electric field through Eq. (\ref{F-theta}).
Now we study the dynamics of $\bm{m}$ and $\bm{n}$ phenomenologically, i.e., based on the Landau-Lifshitz-Gilbert equation  \cite{Hals2011}.
From the total free energy of the system $F_{\rm AF}=F_0+F_\theta$, the effective fields for $\bm{n}$ and $\bm{m}$ are given by
\begin{align}
\begin{split}
\bm{f}_n=-\frac{\delta F_{\rm AF}}{\delta\bm{n}}&=A\bm{n}\times(\nabla^2\bm{n}\times\bm{n})+Kn_z\bm{e}_z-(\bm{n}\cdot\bm{H})\bm{m}\\
&\quad+\eta_\theta(\bm{E}_{\mathrm{ac}}\cdot\bm{B})\bm{e}_{[111]},\\
\bm{f}_m=-\frac{\delta F_{\rm AF}}{\delta\bm{m}}&=-a\bm{m}+\bm{n}\times(\bm{H}\times\bm{n}),
\label{fmn}
\end{split}
\end{align}
where $\eta_\theta=(e^2/2\pi h)(\sqrt{3}Un_0/M_0)$.
Since the AF insulator has the HM contact, in the resonance state a pure spin current is injected into the HM layer through the interface, which enhances the Gilbert damping constant \cite{Cheng2014}.
The Landau-Lifshitz-Gilbert equation is given by
\begin{align}
\begin{split}
\dot{\bm{n}}&=(\gamma\bm{f}_m-G_1\dot{\bm{m}})\times\bm{n},\\
\dot{\bm{m}}&=(\gamma\bm{f}_n-G_2\dot{\bm{n}})\times\bm{n}+(\gamma\bm{f}_m-G_1\dot{\bm{m}})\times\bm{m}+\tau_{\mathrm{SP}},
\label{llg}
\end{split}
\end{align}
where $\gamma=1/\hbar$, $G_1$ and $G_2$ are dimensionless Gilbert-damping parameters, and $\tau_{\mathrm{SP}}=-G_{\mathrm{SP}}(\dot{\bm{n}}\times\bm{n}+\dot{\bm{m}}\times\bm{m})$ is the additional damping torque with a spin pumping parameter $G_{\mathrm{SP}}$ \cite{Cheng2014,Takei2014}. 

To obtain the resonance state, where all the spins are precessing uniformly, we assume the dynamics of the N\'{e}el vector and the total magnetization around the easy axis as $\bm{n}(t)=\bm{e}_z+\delta\bm{n}(t)$ and $\bm{m}(t)=\delta\bm{m}(t)$, denoting that $\delta\bm{n}(t)$ and $\delta\bm{m}(t)$ are the small precession components ($|\delta\bm{n}|, |\delta\bm{m}|\ll1$).
After the linearization and the Fourier transform $\delta\bm{n}(t)=\int \delta\tilde{\bm{n}}(\omega)e^{-i\omega t}d\omega/(2\pi)$, Eq. (\ref{llg}) reduces to
\begin{align}
\begin{split}
2i\omega_H\omega\delta\tilde{\bm{n}}/\omega_a
+\left[ \left(\omega^2+\omega_{H}^2\right)/\omega_{a}-\omega_{K}+i\omega\tilde{G}_2\right]
\bm{e}_z\times\delta\tilde{\bm{n}}\\
=\bm{W}\delta(\omega_{0}-\omega),\label{llgl}
\end{split}
\end{align}
where $\omega_{H}=\gamma g\mu_B B$, $\omega_{a}=\gamma a$, $\omega_{K}=\gamma K$, $\bm{W}=\omega_{\theta}\bm{e}_{[111]}\times\bm{e}_z$ with $\omega_{\theta}=\gamma \eta_\theta E_{\mathrm{ac}}B$, $\omega_{0}$ is the frequency of the applied ac electric field [$\bm{E}_{\mathrm{ac}}(t)=E_{\mathrm{ac}}e^{i\omega_{0}t}\bm{e}_z$], and $\tilde{G}_2=G_2+G_{\mathrm{SP}}$.
It can be shown that for $\tau \gg \hbar/J$ ($\tau$ is the typical time scale of AF dynamics and $J\propto A$ is the AF exchange coupling constant), the $G_1$ term in Eq.~(\ref{llg}) becomes unimportant, enabling
the disregard of the damping term $-G_1\dot{\bm{m}}\times\bm{n}$ \cite{Takei2015}.
The resonance frequencies are obtained as \cite{Comment1}
\begin{align}
\begin{split}
\omega_p=\sqrt{\omega_{a}\omega_{K}}+p\omega_{H},
\end{split}
\end{align}
where $p=+(-)$ corresponds to the excitation of the right-handed (left-handed) mode.
Note that these frequencies do not depend on the parameters of the $\theta$ term.
This is because the $\theta$ term acts only as the driving force to cause the resonance, as is seen from Eq. (\ref{llgl}).

%%%%%%%%%%%
{\it Detection of the antiferromagnetic resonance.---}
So far we have shown that the AF resonance can be realized by ac electric fields, which is in sharp contrast to conventional methods using ac magnetic fields.
How can we detect this electrically driven AF resonance?
Regarding its detection, we can employ a standard method.
Namely we can observe the spin-pumping-induced voltage signal in the HM layer.
One of the advantages of employing this method in our system is that we can identify its detection easily.
Since the system we consider is insulating, we are free from additional dc voltages from the anisotropic magnetoresistance and the anomalous Hall effect \cite{Harder2011}.
As shown in Fig. \ref{Fig1}(b), the spin pumping generates a pure dc spin current $\bm{J}_{s}$ flowing across the AF insulator/HM interface as $\bm{J}_s=(\hbar/e)\Gamma_{\mathrm{eff}}^r\left\langle \bm{n}\times\dot{\bm{n}}\right\rangle_{t}$ with $\left\langle \cdots\right\rangle_{t}$ indicating time average \cite{Chiba2014}.
Here, $\Gamma_{\mathrm{eff}}^r$ is the real part of the effective mixing conductance (reflecting the influence of a back flow spin current) per unit area \cite{Tserkovnyak2005,Chiba2014}.
Its magnitude $J_s=|\bm{J}_s|$ is given by
\begin{align}
\begin{split}
J_s(\omega_{0})
=2\frac{\hbar}{e}\Gamma_{\mathrm{eff}}^r\omega_{0}\operatorname{Im}\left[\delta\tilde{n}_{x}(\omega_0)\delta\tilde{n}_{y}^{*}(\omega_0)\right].\label{J_s}
\end{split}
\end{align}
The spin polarization in the HM decays due to spin relaxation with the length scale characterized by the spin diffusion length \cite{Niimi2015}. Here, we assume that the spin relaxation is included in $\Gamma_{\mathrm{eff}}^r$ of the HM.
The spin current is converted into an electric voltage across the transverse direction via the inverse spin Hall effect \cite{Saitoh2006}: $V_{\mathrm{SP}}(\omega_{0})=\rho d_H\alpha_{\mathrm{SH}}J_{s}(\omega_{0})$, where $d_H$ is the thickness of the HM and $\alpha_{\mathrm{SH}}$ is the spin Hall angle.
Using Eq. (\ref{J_s}), $V_{\mathrm{SP}}(\omega_{0})$ is written explicitly as \cite{comment2}
\begin{align}
\begin{split}
V_{\mathrm{SP}}(\omega_{0})
=&-\frac{1}{8}\rho d_H\alpha_{\mathrm{SH}}\frac{\hbar}{e}\Gamma_{\mathrm{eff}}^r\omega_{0}\\
&\times\sum_{p=\pm}p\frac{\omega_H}{\omega_K}\sqrt{\frac{\omega_a}{\omega_K}}\frac{\omega_\theta^2\sin^2\theta_{[111]}}{(g_2^p\omega_0)^2}\operatorname{Lor}(\omega_{0},\omega_{p})
,\label{V_SP}
\end{split}
\end{align}
where $\operatorname{Lor}(\omega_{0},\omega_{p})=(g_2^p\omega_0)^2/[(\omega_{0}-\omega_{p})^2+(g_2^p\omega_0)^2]$ is a symmetric spectrum function (Lorentzian), $g_2^p$ is a constant, and $\theta_{[111]}$ being the angle between $\bm{e}_{[111]}$ and $\bm{e}_z$.
For example, in the case of $B=0.1\ \mathrm{T}$ and $E_{\mathrm{ac}}=1\ \mathrm{V/m}$ with possible (typical) values of the parameters, we find the magnitude of $V_{\mathrm{SP}}$ in the resonance state as $V_{\mathrm{SP}}(\omega_\pm)\sim 10\ \mu\mathrm{V}$ \cite{comment2}.
This value is experimentally observable.
Furthermore, it should be noted that the above value of the ac electric field, $E_{\mathrm{ac}}=1\ \mathrm{V/m}$, is small.
Namely, from the viewpoint of lower energy consumption, our proposal has an advantage compared to conventional ``current-induced'' methods that require such high-density currents as $\sim 10^{10}\ \mathrm{A/m^2}$ \cite{Brataas2012}.

One can confirm the electric-field-induced AF resonance in this study as follows.
The experimental setup we propose is for measuring the magnetic-field angle dependence of the induced dc voltage in the case where the applied static magnetic field $\bm{B}$ is much weaker than both the AF exchange coupling and uniaxial anisotropy.
In this case, the induced voltage $V_{\mathrm{SP}}(\omega_\pm)\propto-(\gamma \eta_\theta E_{\mathrm{ac}}B)^2\cos^2\psi$ oscillates as a function of the relative angle $\psi$ between $\bm{E}_{\mathrm{ac}}$ and $\bm{B}$.
For magnetic field rotations in the $x$-$y$ plane, where $\bm{E}_{\mathrm{ac}}\perp\bm{B}$, one will not find any electrical signals since $\psi=\pi/2$.
Such a $\psi$ dependence is a direct consequence of the existence of the topological term (\ref{F-theta}) which is essential for the realization of the electric-field-induced AF resonance in this study.

%%%%%%%%%%%
{\it Discussions.---}
First, we comment on a possible realization of our prediction in real materials.
It is suggested theoretically that the AF insulator phases, in which there exists the $\theta$ term and the value of $\theta$ is proportional to the N\'{e}el field as in our case, can be realized in the magnetically doped Bi$_2$Se$_3$ family \cite{Li2010} and transition metal oxides with the corundum structure \cite{Wang2011}.
Recently, AF insulator phases have been observed experimentally in Ga$_x$Bi$_{2-x}$Se$_3$ \cite{Kim2015} and Ce$_x$Bi$_{2-x}$Se$_3$ \cite{Lee2015}.
These could be candidate materials to observe the electric-field-induced AF resonance.

Next, we discuss the mechanism of the AF resonance in this study.
Recall that the presence of the $\theta$ term (\ref{theta-term}) results in the topological magnetoelectric effect (in ground states), i.e., an electric polarization density in the bulk is obtained as $\bm{P}=(e^2/2\pi h)\theta\bm{B}$ \cite{Hasan2010,Qi2011,Ando2013,Qi2008}.
In the case of static magnetic fields, the time derivative of both sides reads $\dot{\bm{P}}=(e^2/2\pi h)\dot{\theta}\bm{B}$.
There is no electric-field screening since the system we consider is insulating, which means that the electric polarization in the bulk can be flipped by external ac electric fields.
Namely, we have demonstrated that nonzero $\dot{\bm{P}}$ realized by external ac electric fields induces nonzero $\dot{\theta}$, i.e., a time dependence of the N\'{e}el field $\bm{n}$.
On the other hand, a recent study has proposed a novel phenomenon, the ``dynamical chiral magnetic effect'' in AF insulators with SOC that possess the $\theta$ term \cite{Sekine2016}.
The dynamical chiral magnetic effect indicates an alternating electric current generation by magnetic fields such that $\bm{j}=(e^2/2\pi h)\dot{\theta}\bm{B}$, and emerges as a consequence of the realization of the dynamical axion field $\theta(t)$ in condensed matter.
Here, nonzero $\dot{\theta}$, i.e., a time dependence of the N\'{e}el field $\bm{n}$ realized by external ac magnetic fields, induces a polarization current $\bm{j}(=\dot{\bm{P}})$ in the bulk.
Therefore, the electric-field-induced AF resonance in this study can be understood as the \textit{inverse process} of the dynamical chiral magnetic effect.
In other words, we have proposed a way to electrically induce the dynamical axion field in condensed matter.

\begin{figure}[!t]
\centering
\includegraphics[width=0.85\columnwidth]{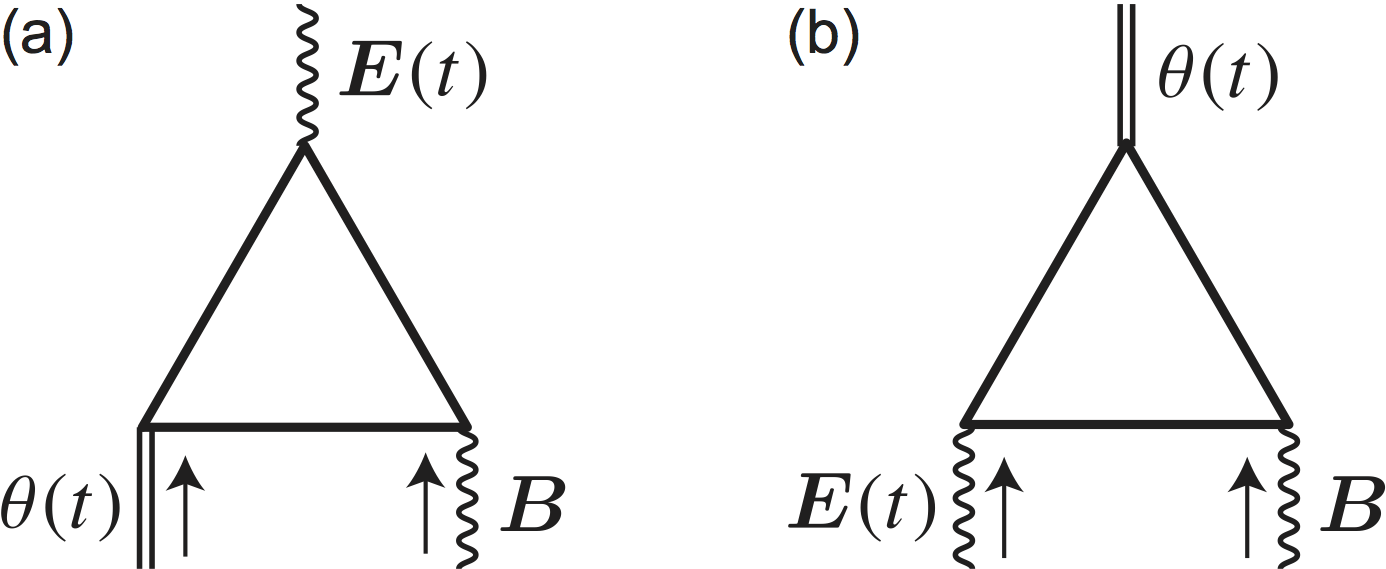}
\caption{(a) Diagrammatic representation of the dynamical chiral magnetic effect.
$\bm{E}(t)[\propto \bm{j}(t)]$ represents an alternating current induced by external $\theta(t)$ and $\bm{B}$.
(b) That of the inverse process of the dynamical chiral magnetic effect.
$\theta(t)$ represents an AF resonance state induced by external $\bm{E}(t)$ and $\bm{B}$.
These two phenomena are described by the $\theta$ term $S_\theta=(e^2/2\pi h)\int dt d^3r \theta(t)\bm{E}\cdot\bm{B}$ with $\theta(t)\propto\sum_f n_f(t)$.
Arrows indicate the inputs induced by external forces.
Solid lines indicate the non-interacting Green's function of the electrons.
}\label{Fig2}
\end{figure}
We can also confirm this mechanism from another viewpoint.
As shown in Fig. \ref{Fig2}, it is known that the $\theta$ term represents the chiral anomaly described by a triangle Feynman diagram.
In this triangle diagram, one of the three vertices is the N\'{e}el field $i\gamma^5 M_{5f}$ and the other two are the electromagnetic fields $e\gamma^\mu A_\mu$ \cite{Sekine2016}.
Both the AF resonance in this study and the dynamical chiral magnetic effect are described by the $\theta$ term.
Figure \ref{Fig2} shows a diagrammatic comparison of these two phenomena.
In the dynamical chiral magnetic effect, the N\'{e}el field and the magnetic field are the inputs, and the electric field is the observable.
On the other hand, in the AF resonance state, the magnetic and electric fields are the inputs, and the N\'{e}el field is the observable.
Hence, we see that the AF resonance state can be understood as the inverse process of the dynamical chiral magnetic effect.

%%%%%%%%%%%
{\it Summary.---}
In summary, we have demonstrated that the AF resonance can be realized by ac electric fields in 3D AF insulators with SOC.
This is in sharp contrast to conventional methods using ac magnetic fields.
It is found that weak ac electric fields $\sim 1\ \mathrm{V/m}$ are enough to cause the resonance.
The essential point is the existence of the $\theta$ term which arises as a consequence of strong SOC.
The mechanism of the AF resonance in this study can be understood as the inverse process of the dynamical chiral magnetic effect.
In other words, we have proposed a way to electrically induce the dynamical axion field in condensed matter.
Also, the observation of the electric-field-induced AF resonance indicates the existence of the chiral magnetic effect.
The spin-pumping-induced voltage signal via the inverse spin Hall effect, which is an observable quantity to verify our prediction, is characterized by the angle dependence between the applied ac electric field and static magnetic field.
Our study opens a new direction in possible applications of topological materials in spintronics.

\vspace{1ex}
The authors thank Y. Araki, S. Takahashi, K. Nomura, and G. E. W. Bauer for fruitful discussions.
The authors are supported by JSPS Research Fellowships.

%\appendix

\begin{widetext}
\vspace{2ex}
\begin{center}
\textbf{{\large Supplemental Material}}
\end{center}
\vspace{2ex}
\textbf{1. Solution of Antiferromagnetic Dynamics}
\vspace{2ex}

Here we derive the solution of the Landau-Lifshitz-Gilbert equation (\ref{llgl}) in the resonance state.
Equation (\ref{llgl}) is rewritten in the matrix form
\begin{align}
\begin{split}
\begin{bmatrix}
2i\omega\omega_{H}
& -\left(  \omega^2-\omega_{a}\omega_{K}+\omega_{H}^2+i\omega\omega_{a}\tilde{G}_2\right) \\
\omega^2-\omega_{a}\omega_{K}+\omega_{H}^2+i\omega\omega_{a}\tilde{G}_2
& 2i\omega\omega_{H}
\end{bmatrix}
\begin{bmatrix}
\delta \tilde{n}_{x}(\omega) \\ \delta \tilde{n}_{y}(\omega)
\end{bmatrix}
=\omega_{a}\delta(\omega_{0}-\omega)
\begin{bmatrix}
W_x \\ W_y
\end{bmatrix},
\end{split}\tag{S1}
\end{align}
where $\omega_{H}=\gamma g\mu_B B$, $\omega_{a}=\gamma a$, $\omega_{K}=\gamma K$, $\bm{W}=\omega_{\theta}\bm{e}_{[111]}\times\bm{e}_z$ with $\omega_{\theta}=\gamma \eta_\theta E_{\mathrm{ac}}B$, $\omega_{0}$ is the frequency of the applied ac electric field [$\bm{E}_{\mathrm{ac}}(t)=E_{\mathrm{ac}}e^{i\omega_{0}t}\bm{e}_z$], and $\tilde{G}_2=G_2+G_{\mathrm{SP}}$. 

As in the case of ferromagnets \cite{FMR-Book}, we multiply the inverse matrix from the left hand side.
The final form reads
\begin{align}
\begin{split}
\begin{bmatrix}
\delta \tilde{n}_{x}(\omega) \\ \delta \tilde{n}_{y}(\omega)
\end{bmatrix}
=\ 
\begin{bmatrix}
\chi_1(\omega)
&  \chi_2(\omega) \\
-\chi_2(\omega)
& \chi_1(\omega)
\end{bmatrix}
\begin{bmatrix}
W_x \\ W_y
\end{bmatrix},
\label{matrix}
\end{split}\tag{S2}
\end{align}
where the susceptibility is defined as
\begin{align}
\begin{split}
\begin{bmatrix}
\chi_1(\omega)
&  \chi_2(\omega) \\
-\chi_2(\omega)
& \chi_1(\omega)
\end{bmatrix}
=&\ 
\frac{
\omega_{a}\delta(\omega_{0}-\omega)
}{(\omega^2-\omega_+^2)(\omega^2-\omega_-^2)-2i\omega(\omega_+\omega_-\omega_a\tilde{G}_2)}
\\
&\times
\begin{bmatrix}
2i\omega\omega_{H}
&  \omega^2-\omega_{a}\omega_{K}+\omega_{H}^2+i\omega\omega_{a}\tilde{G}_2 \\
-\left( \omega^2-\omega_{a}\omega_{K}+\omega_{H}^2+i\omega\omega_{a}\tilde{G}_2\right)
& 2i\omega\omega_{H}
\end{bmatrix}
\label{susceptibility}
\end{split}\tag{S3}
\end{align}
with $\omega_{\pm}=\sqrt{\omega_{a}\omega_{K}}\pm\omega_{H}$ the resonance frequencies.

\vspace{6ex}
\noindent\textbf{2. Estimation of the magnitude of the voltage $\bm{V_{\mathrm{SP}}}$}
\vspace{2ex}

In order to estimate the magnitude of the voltage $V_{\mathrm{SP}}$, we first need to obtain an explicit form of the spin current $\bm{J}_s$ induced by the spin pumping.
Applying the resonance approximation, the magnitude of the spin current $J_s=|\bm{J}_s|$ generated through the antiferromagnetic (AF) resonance is given by
\begin{align}
\begin{split}
J_s(\omega_{0})
&=2\frac{\hbar}{e}\Gamma_{\mathrm{eff}}^r\omega_{0}\operatorname{Im}\left[  \delta\tilde{n}_{x}\delta\tilde{n}_{y}^{*}\right]\\
&=2\frac{\hbar}{e}\Gamma_{\mathrm{eff}}^r\omega_{0}\left(  -\operatorname{Im}\left[  \chi_1\chi_2^{*}\right]W_x^2+\operatorname{Im}\left[  \chi_2\chi_1^{*}\right]W_y^2\right)\\
&=-2\frac{\hbar}{e}\Gamma_{\mathrm{eff}}^r\omega_{0}\frac{
2\omega_{a}^2\omega_H\omega_{0}(\omega_{0}^2-\omega_+\omega_-)
}{(\omega_{0}^2-\omega_+^2)^2(\omega_{0}^2-\omega_-^2)^2+\left(  2\omega_+\omega_-\omega_a\tilde{G}_2\right)^2\omega_{0}^2}\left(  W_x^2+W_y^2\right)\\
&\to -\frac{\hbar}{2e}\Gamma_{\mathrm{eff}}^r\omega_{0}\sum_{p=\pm}p\frac{\omega_H}{\omega_K}\sqrt{\frac{\omega_a}{\omega_K}}\frac{W_x^2+W_y^2}{(g_2^p\omega_{0})^2}\operatorname{Lor}(\omega_{0},\omega_{p})\ \ \ (\omega_{0}\to\omega_{\pm}),
\end{split}\tag{S4}
\end{align}
where $\operatorname{Lor}(\omega_{0},\omega_{p})=(g_2^p\omega_{0})^2/[(\omega_{0}-\omega_{p})^2+(g_2^p\omega_{0})^2]$ describes the symmetric spectrum function (Lorentzian) with $\omega_{p}=\sqrt{\omega_{a}\omega_{K}}+p\omega_{H}$, and $g_2^p=\omega_+\omega_-\omega_a\tilde{G}_2/[\omega_p(\omega_++\omega_-)\sqrt{\omega_+^2-\omega_-^2}]$.
Here,
\begin{align}
\begin{split}
\Gamma_{\mathrm{eff}}^r
=\frac{1}{\rho d_H}\frac{2\lambda\rho \Gamma^{r}\tanh\frac{d_{H}}{2\lambda}}{1+2\lambda\rho \Gamma^{r}\coth\frac{d_{H}}{\lambda}}
\end{split}\tag{S5}
\end{align}
is the real part of the effective mixing conductance (reflecting the influence of a back flow spin current) per unit area \cite{Tserkovnyak2005,Chiba2014}, where $\rho$ is the resistivity of the heavy metal (HM), $d_H$ the thickness of the HM, $\lambda$ the spin diffusion length of the HM, and $\Gamma^r$ the real part of the mixing conductance at the AF insulator/HM interface.
With the use of the relations $W_x^2+W_y^2=\frac{1}{4}\omega_\theta^2\sin^2\theta_{[111]}$ and $V_{\mathrm{SP}}(\omega_{0})=\rho d_H\alpha_{\mathrm{SH}}J_{s}(\omega_{0})$, we arrive at Eq. (\ref{V_SP}):
\begin{align}
\begin{split}
V_{\mathrm{SP}}(\omega_{0})
=-\frac{1}{8}\rho d_H\alpha_{\mathrm{SH}}\frac{\hbar}{e}\Gamma_{\mathrm{eff}}^r\omega_{0}\sum_{p=\pm}p\frac{\omega_H}{\omega_K}\sqrt{\frac{\omega_a}{\omega_K}}\frac{\omega_\theta^2\sin^2\theta_{[111]}}{(g_2^p\omega_0)^2}\operatorname{Lor}(\omega_{0},\omega_{p}).\label{SM-V_SP}
\end{split}\tag{S6}
\end{align}

Let us estimate the magnitude of the voltage $V_{\mathrm{SP}}$ in the resonance state ($\omega_0=\omega_\pm$).
As a possible case, we set $B=0.1\ \mathrm{T}$ and $E_{\mathrm{ac}}=1\ \mathrm{V/m}$.
We consider an AF insulator of $1\ \mu\mathrm{m^3}$ attached with a HM (Pt) of $d_H=10\ \mathrm{nm}$.
In Pt, we get $\rho= 4.1\times 10^{-7}\ \Omega\cdot\mathrm{m}$, $\lambda= 1.4\ \mathrm{nm}$, and $\alpha_{\mathrm{SH}}= 0.12$ \cite{Obstbaum2014}.
We use typical values for antiferromagnets such that $a\approx J$ and $K\sim 0.01J$ with $J\sim 1\ \mathrm{meV}$ (which leads to $\sqrt{\omega_a\omega_K}\sim 100\ \mathrm{GHz}$), and assume that $\sin^2\theta_{[111]}\sim 0.1$.
Also, we use possible values at the AF insulator/HM interface such that $\Gamma^r\sim 10^{14}\ \Omega^{-1}\mathrm{m}^{-2}$ and $G_2(\sim G_{\mathrm{SP}})\sim 10^{-3}$ \cite{Takei2014}.
In the $\theta$ term (\ref{theta-term}), we have retained only the leading term, i.e., $Un_0/M_0\sim 0.1$ (which leads to $\omega_\theta\sim 1\ \mathrm{GHz}$) \cite{Sekine2016}.
Substituting these possible (typical) parameter values into Eq. (\ref{SM-V_SP}), we obtain $V_{\mathrm{SP}}(\omega_\pm)\sim 10\ \mu\mathrm{V}$.

\end{widetext}

\end{document}